\documentclass[preprint,preprintnumbers,amsmath,amssymb]{revtex4}
\usepackage{graphicx}
\newcommand{\vlk}{$V_{{\rm low}-k}$ }
\newcommand{\vlkn}{$V_{{\rm low}-k}$}

\begin{document}
%\topmargin=0.1cm
%\bottommargin=0.1cm
\title{Nuclear matter with Brown-Rho-scaled \\ Fermi liquid interactions}

\author{Jeremy W.\ Holt, G.\ E.\ Brown, Jason D.\ Holt, and T.\ T.\
  S.\ Kuo}
\affiliation{Department of Physics, SUNY, Stony Brook, New York 11794, USA}

\date{\today}

\begin{abstract}
We present a description of symmetric nuclear matter within the framework of
Landau Fermi liquid theory. The low momentum nucleon-nucleon interaction \vlk
is used to calculate the effective interaction between quasiparticles on the
Fermi surface, from which we extract the quasiparticle effective
mass, the nuclear compression modulus, the symmetry energy, and the anomalous
orbital gyromagnetic ratio. The exchange of density, spin, and isospin
collective excitations is included through the Babu-Brown induced
interaction, and it is found that in the absence of three-body forces the
self-consistent solution to the 
Babu-Brown equations is in poor agreement with the empirical values for the
nuclear observables. This is improved by lowering the nucleon and meson masses
according to Brown-Rho scaling, essentially by including a scalar tadpole
contribution to the meson and nucleon masses, as well as by scaling
$g_A$. We suggest that modifying the masses of the exchanged mesons is
equivalent to introducing a short-range three-body force, and the net result
is that the Brown-Rho double decimation \cite{brown5} is accomplished all at
once.
\end{abstract}

\maketitle

\section{Introduction}
Landau's theory of normal Fermi liquids \cite{landau1,landau2,landau3}
describes strongly interacting many-body systems in terms of weakly interacting
quasiparticles. Provided that the quasiparticles lie sufficiently close to the 
Fermi surface, they will be long-lived and constitute appropriate degrees of
freedom for the system. The central aim of the theory is to
determine the quasiparticle interaction, either phenomenologically or
microscopically, with which it is possible to describe the low-energy,
long-wavelength excitations of the system. This, in turn, is sufficient for the
description of many bulk equilibrium properties of the interacting Fermi
system. The initial application of Fermi liquid theory to nuclear physics was
the phenomenological description of finite nuclei and nuclear matter by Migdal
\cite{migdal1,migdal2}, and later a microscopic approach to Fermi liquid
theory based on the Brueckner-Bethe-Goldstone reaction matrix theory was
developed by B\"{a}ckman \cite{backman} and others \cite{brown1,poggioli} to
describe nuclear matter. Although the latter approach was quantitatively
successful, it was observed \cite{poggioli} that Brueckner-Bethe-Goldstone
theory is less reliable in the vicinity of the Fermi surface due to the use of
angle-averaged Pauli operators and the unsymmetrical treatment of particle and
hole self energies, which leads to an unphysical energy gap at the Fermi
surface.

With the recent development of a nearly universal low-momentum nucleon-nucleon
(NN) interaction \vlk \cite{bogner1} derived from renormalization group
methods, the application of Fermi liquid theory to nuclear matter has received
renewed attention \cite{schwenk1,schwenk2,schwenk3}. The strong 
short-distance repulsion incorporated into all high-precision NN potential
models is integrated out in low momentum interactions, rendering them suitable
for perturbation theory calculations. Although limited Brueckner-Hartree-Fock
studies 
\cite{kukei} indicate that saturation is not achieved with \vlk at a fixed
momentum cutoff $\Lambda$, it has recently been shown \cite{bogner2} that by
supplementing \vlk with the leading-order chiral three-nucleon force, nuclear
matter does saturate, thereby justifying the use of \vlk in studies of nuclear
matter.

Although many properties of the interacting ground state are beyond the scope
of Fermi liquid theory, the quasiparticle interaction is directly related to
several nuclear observables, including the compression modulus, symmetry
energy, and anomalous orbital gyromagnetic ratio. As 
originally shown by Landau, the quasiparticle interaction is obtained from a
certain limit of the four-point vertex function in the particle-hole
channel. It is well known that using realistic NN interactions in the lowest
order approximation to the quasiparticle interaction is insufficient to
stabilize nuclear matter, as evidenced by a negative value of the compression
modulus. This general phenomenon is observed in our calculations with \vlk as
well. However, stability is achieved 
by treating the exchange of density, spin, and isospin collective excitations
to all orders in perturbation theory. The inclusion of these virtual
collective modes in the quasiparticle interaction is carried out through the
induced interaction formalism of Babu and Brown \cite{babu}, which was
originally developed for the description of liquid $^3$He and later applied to
nuclear matter by Sj\"{o}berg \cite{sjoberg1,sjoberg2}. Subsequent work
\cite{dickhoff,backman2} has confirmed the importance of the induced
interaction in building up correlations around a single quasiparticle, thereby
increasing the compression modulus.

Our study is motivated in part by the work of Schwenk {\it et al.}
\cite{schwenk1}, who were able to predict the spin-dependent parameters of the 
quasiparticle interaction from the experimentally extracted spin-independent
parameters. Crucial to these calculations was a novel set of sum rules, derived
from the induced interaction formalism, based on a similar treatment by Bedell
and Ainsworth \cite{bedell} to liquid $^3$He. In this paper we present a fully
self-consistent solution to the Babu-Brown induced interaction equations for
symmetric nuclear matter. Our iterative solution turns out to be qualitatively
similar to the results of \cite{schwenk1}, but we find that at nuclear matter
density the compression modulus and symmetry energy are smaller than the
experimentally observed values while the anomalous orbital gyromagnetic ratio
is too large, suggesting the possibility that important phenomena have been
neglected.

We propose to extend this study by including hadronic
modifications associated with the partial restoration of chiral symmetry at
nuclear matter density, as suggested in \cite{brownrho}. In this scenario,
referred to as Brown-Rho scaling, the dynamically generated hadronic masses
drop in the approach to chiral restoration, and at nuclear matter density it
is expected that the masses of the light hadrons (other than the masses of the
pseudoscalar mesons, which are protected by their Goldstone nature) decrease by
approximately 20\%. The success of one-boson-exchange and chiral EFT
potentials in describing 
the nucleon-nucleon interaction suggests that a modification of meson masses
in medium ought to have verifiable consequences in low energy nuclear
physics. Although there is much current theoretical and experimental effort
devoted to the program of assessing these medium modifications, the
consequences for low-energy nuclear physics have yet to be fully explored.

Applying the mass scaling suggested in \cite{brownrho} to our calculations of
nuclear matter, we obtain a set of Fermi liquid coefficients in better
agreement with both experiment and the nontrivial sum rules derived in
\cite{schwenk1}. Explicit three-body forces, though essential for a complete
description of nuclear matter, have been neglected in this study. However, we
argue that modifying the vector meson masses is equivalent to including a
specific short-ranged three-body force. We conclude with a discussion of the
consequences of Brown-Rho scaling on the tensor force, which is diminished by
the increasing strength of $\omega$-meson exchange.

%One of the fundamental issues in contemporary nuclear physics is the
%description of hadronic properties as the temperature or density increases
%toward chiral restoration. It has been suggested that even at nuclear matter
%density hadronic masses and widths will be modified due to the partial
%restoration of chiral symmetry \cite{brownrho}*. The success of
%one-boson-exchange potentials in describing the nucleon-nucleon interaction
%suggests that a modification of meson masses in medium ought to have
%verifiable consequences in low energy nuclear physics. The consequences for 
%low-energy nuclear physics, though surely significant, have yet to be fully
%explored. In this paper we wish to consider the role of in-medium hadronic
%modifications to the properties of nuclear matter within the formalism of
%Landau Fermi liquid theory.

\section{Fermi liquid theory}

In this section we present a short description of Fermi liquid
theory and its application to nuclear physics with emphasis on the microscopic
foundation of the theory. The main assumption underlying 
Landau's description of many-body Fermi systems is that there is a one-to-one
correspondence between states of the ideal system and states 
of the interacting system. As one gradually turns on the interaction, the
noninteracting particles become ``dressed'' through interactions with the
many-body medium and evolve into weakly interacting quasiparticles. The
interacting system is in many ways similar to an ideal system 
in that the classification of energy states remains unchanged and there is a
well-defined Fermi surface, but the quasiparticles acquire an effective
mass $m^*$ and finite lifetimes \mbox{$\tau \sim (k-k_F)^{-2}$}. The energy of
the interacting system is a complicated functional of the quasiparticle
distribution function, and in general the exact dependence 
is inaccessible. But one can extract important information about bulk
properties of the system by considering small changes in the distribution
function. Expanding to second order, one finds
\begin{equation}
\delta E = \sum_{{\bf k}_1} \epsilon_{{\bf k}_1}^{(0)} \delta
n({\bf k}_1)+\frac{1}{2\Omega}\sum_{{\bf k}_1,{\bf k}_2} f({\bf k}_1,{\bf
  k}_2) 
\delta n({\bf k}_1) \delta n({\bf k}_2) + {\cal O}(\delta n^3).
\end{equation}
In this equation $\Omega$ is the volume of the system, $\epsilon_{{\bf
k}_1}^{(0)}$ is the energy added to the system by introducing a single
quasiparticle with momentum ${\bf k}_1$ (note that for $|{\bf k}_1| \equiv
k_1=k_F$, $\epsilon_{{\bf k}_1}^{(0)}$ is just the chemical potential), and
$f({\bf k}_1,{\bf k}_2)$ describes the interaction between two quasiparticles.

Since the quasiparticle interaction $f({\bf k}_1,{\bf k}_2)$ is the
fundamental quantity of interest in Fermi liquid theory, we will carefully
discuss its properties and its relationship to nuclear observables. Assuming
the interaction to be purely exchange, it can be written as
\begin{equation}
f({\bf k}_1,{\bf k}_2)=\frac{1}{N_0}[
F({\bf k}_1,{\bf k}_2) + F^\prime({\bf k}_1,{\bf k}_2) \tau_1 \cdot
\tau_2 + G({\bf k}_1,{\bf k}_2)\sigma_1 \cdot \sigma_2 + G^\prime({\bf
  k}_1,{\bf k}_2) \tau_1 \cdot \tau_2 \sigma_1 \cdot \sigma_2],
\label{ffunction}
\end{equation}
where we have factored out the density of states per unit volume at the Fermi
surface, $N_0=\frac{2m^*k_F}{\hbar^2\pi^2}$, which leaves dimensionless Fermi
liquid parameters denoted by $F,G,F^\prime,G^\prime$. The spin-orbit
interaction is neglected because it vanishes in the long wavelength limit in
which we will be interested. Also, we have not included tensor 
operators (which would greatly complicate our calculation) because the tensor
force contributes almost completely in second order, as shown in the original
paper by Kuo and Brown \cite{kb}, as an effective central interaction in the
$^3S_1$ state. In \cite{schwenk1} the tensor Fermi liquid parameters for
symmetric nuclear matter were calculated from \vlk in which the dominant
second-order contributions from one-pion exchange were
included. Since quasiparticles are well-defined only near the Fermi 
surface, we assume that $k_1 = k_F = k_2$. In this case the
dimensionless Fermi liquid parameters $F,F^\prime, G, G^\prime$
depend on only the angle between ${\bf k}_1$ and ${\bf k}_2$, which we call
$\theta$. Then it is convenient to perform a Legendre polynomial expansion as
follows 
\begin{equation}
F({\bf k},{\bf k}^\prime) = \sum_l F_l P_l(\mbox{cos } \theta), \hspace{.1in}
G({\bf k},{\bf k}^\prime) = \sum_l G_l P_l(\mbox{cos } \theta), \mbox{ etc.}
\end{equation}
The Fermi liquid parameters $F_l, G_l, \dots$ decrease rapidly for
larger $l$, and so there are only a small number of parameters that can either
be fit to experiment or calculated microscopically.

In the original application of the theory to liquid $^3$He and nuclear
systems, the quasiparticle interaction was obtained phenomenologically by
fitting the dimensionless Fermi liquid parameters to relevant
data. For nuclear matter several important relationships exist between nuclear
observables and the Fermi liquid parameters. Galilean invariance can be used
\cite{landau1} to connect the Landau parameter $F_1$ to the quasiparticle
effective mass 
\begin{equation}
\frac{m^*}{m}=1+F_1/3.
\label{effmass}
\end{equation}
Adding a small number of neutrons and removing the same number of
protons from the system will increase and decrease, respectively, the density
of protons and neutrons in the system (and therefore the Fermi energies of the
two species). The change in the energy, described by the symmetry energy
$\beta$, can be related \cite{migdal2} to the Landau parameter $F_0^\prime$
\begin{equation}
\beta = \frac{\hbar^2k_F^2}{6m^*}(1+F_0^\prime).
\end{equation}
In a similar way, the equal increase or decrease of the proton and neutron
densities leads to a relationship between the scalar-isoscalar Landau
parameter $F_0$ and the compression modulus ${\cal K}$
\begin{equation}
{\cal K}=\frac{3\hbar^2k_F^2}{m^*}\left (1+F_0\right ).
\end{equation}
Finally, it can be shown \cite{migdal2} that an odd nucleon added just above
the Fermi sea induces a polarization of the medium leading to an anomalous
contribution to the orbital gyromagnetic ratio of the form
\begin{eqnarray}
g_l^p &=& [1 - \delta g_l] \mu_N \nonumber \\
g_l^n &=& [\delta g_l] \mu_N,
\end{eqnarray}
where $\delta g_l$ is given by 
\begin{equation}
\delta g_l = \frac{1}{6}\frac{F^\prime_1 - F_1}{1+F_1/3}.
\end{equation}

Clearly there are certain values of the Landau parameters that are physically
unreasonable. For instance, if $F_1 < -3$ or $F_0 < -1$, the effective
mass or compression modulus would be negative. Quite generally it can be shown
\cite{pomeranchuk} that the Landau parameters must satisfy stability conditions
\begin{equation}
X_l>-(2l+1),
\label{stab}
\end{equation}
where $X$ represents $F,G,F^\prime, G^\prime$.

\begin{figure}
\includegraphics[height=4.5cm]{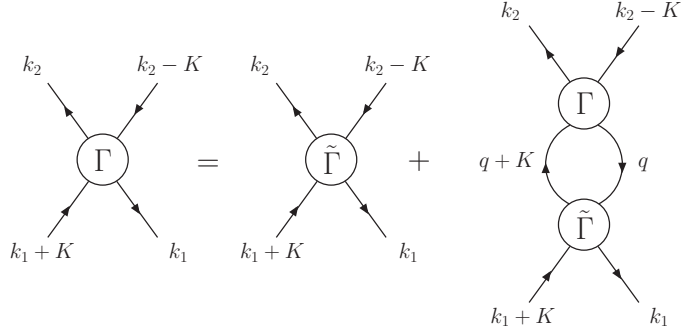}
\caption{The Bethe-Salpeter equation for the fully irreducible vertex function
  $\Gamma$ in terms of the $ph$ irreducible vertex function $\tilde{\Gamma}$.} 
\label{gam}
\end{figure}

A rigorous foundation for the assumptions underlying Landau's theory can be
obtained through formal many-body techniques
\cite{abrikosov, nozieres}. It is not our goal to reproduce the original
arguments \cite{landau3}, but rather to give a clear motivation for the
diagrammatic expansion leading to the quasiparticle interaction. Starting from
the usual definition of the four-point Green's function in momentum space
\begin{eqnarray}
&& G_{\alpha \beta,\gamma \delta}(k_1,k_2;k_3,k_4) = (2\pi)^8
  \delta^{(4)}(k_1+k_2-k_3-k_4)[G_{\alpha 
  \gamma}(k_1) G_{\beta \delta}(k_2) \delta^{(4)}(k_1-k_3) \\ \nonumber
&& - G_{\alpha
  \delta}(k_1) G_{\beta \gamma}(k_2) \delta^{(4)}(k_2-k_3)
 + \frac{i}{(2\pi)^4}G(k_1)G(k_2)G(k_3)G(k_4)\Gamma_{\alpha \beta,\gamma 
  \delta}(k_1,k_2;k_3,k_4)],
\end{eqnarray}
where $G(k_1)$ is the Fourier transform of $G(xt, x^\prime t^\prime)$ and
$k_1,\dots,k_4$ represent four-vectors (e.g.\ $k_1=({\bf k}_1,\omega_1))$, it
can be shown that the quasiparticle interaction is related to a certain 
limit of the four-point vertex function $\Gamma_{\alpha \beta,\gamma
  \delta}(k_1,k_2;k_3,k_4)$. From
energy-momentum conservation ($k_1+k_2=k_3+k_4$) we can write $k_3-k_1 = K =
k_2-k_4$ and therefore define $\Gamma(k_1,k_2;K)= \Gamma(k_1,k_2;k_3,k_4)$. The
important point is that since we are considering only low-energy
long-wavelength excitations, the particle-hole energy-momentum $K$ should be
small. We can write a Bethe-Salpeter equation for the fully reducible vertex
function $\Gamma$ in terms of the {\it ph} irreducible vertex function
$\tilde{\Gamma}$ in the direct channel with momentum transfer $K$:
\begin{eqnarray}
\Gamma_{\alpha \beta,\gamma \delta} (k_1,k_2;K) &=& \tilde{\Gamma}_{\alpha
  \beta, \gamma \delta}(k_1,k_2;K) \nonumber \\
&& -i\sum_{\epsilon, \eta} \int\frac{d^4q}{(2\pi)^4} 
\tilde{\Gamma}_{\alpha \epsilon, \gamma \eta} (k_1,q;K) G(q)G(q+K) \Gamma_{\eta
  \beta, \epsilon \delta}(q,k_2;K)
\end{eqnarray}
shown diagrammatically in Fig.\ \ref{gam}.
The product of propagators may have singularities in the limit that
$K\rightarrow 0$, in which case the poles can be replaced by $\delta$-functions
inside the integral:
\begin{equation}
G(q)G(q+K)=\frac{2i\pi z^2\hat{\bf q}\cdot {\bf K}}{\omega-v_F\hat{\bf q}\cdot
  {\bf K}} 
\delta(\epsilon-\mu)\delta(q-k_F)+\phi({\bf q}),
\label{prop}
\end{equation}
where $z$ is the renormalization at the quasiparticle pole and $\phi({\bf q})$
accounts for the multipair background. The limit $K=(\omega, {\bf K}) \to 0$
depends on the relative ordering of the 
two limits ${\bf K} \rightarrow 0$ and $\omega \rightarrow 0$. Defining
\begin{eqnarray}
\Gamma^\omega(k_1,k_2) &=& \lim_{\omega \rightarrow 0} \lim_{{\bf
    K}\rightarrow 0} 
\Gamma(k_1,k_2;K) \hspace{.1in} \mbox{ and} \nonumber \\
\Gamma^K(k_1,k_2) &=& \lim_{{\bf K} \rightarrow 0} \lim_{\omega \rightarrow 0}
\Gamma(k_1,k_2;K),
\end{eqnarray}
from eq.\ (\ref{prop}) we see that the product of propagators is regular for
$\Gamma^\omega$. Thus, to calculate $\Gamma^\omega$ we must first calculate
the $ph$ irreducible diagrams belonging to $\tilde{\Gamma}$ and then iterate
via the Bethe-Salpeter equation with the intermediate multipair background
$\phi$. The $\delta$-function singularities in $\Gamma^K$ can be used to
perform the integrals over $q_0$ and $|{\bf q}|$, and through algebraic
manipulation it is possible to combine $\Gamma^\omega$ and $\Gamma^K$ into a
single integral equation
\begin{eqnarray}
\Gamma^K_{\alpha \beta,\gamma \delta} (k_1,k_2) &=& \Gamma^\omega_{\alpha
  \beta, \gamma \delta}(k_1,k_2) \nonumber \\
&& -\frac{1}{16\pi}N_0z^2\sum_{\epsilon, \eta} \int d\Omega_q 
\Gamma^\omega_{\alpha \epsilon, \gamma \eta} (k_1,q) \Gamma^K_{\eta
  \beta, \epsilon \delta}(q,k_2).
\label{scat}
\end{eqnarray}
Physically, $\Gamma^\omega$ represents the exchange of virtual excitations
between quasiparticles, and $\Gamma^K$ represents the forward scattering of
quasiparticles at the Fermi surface. By relating these vertex functions to
the equations describing zero sound, Landau \cite{landau3} was able to make the
identifications
\begin{eqnarray}
f(k_1,k_2)&=&z^2\Gamma^\omega(k_1,k_2) \hspace{.1in} \mbox{ and} \nonumber \\
a(k_1,k_2)& =& z^2\Gamma^k(k_1,k_2),
\label{fA}
\end{eqnarray}
where $f(k_1,k_2)$ is just the quasiparticle interaction introduced earlier
and $a(k_1,k_2)$ is the physical scattering amplitude.

\section{Induced Interaction}
\begin{figure}
\includegraphics[height=4cm]{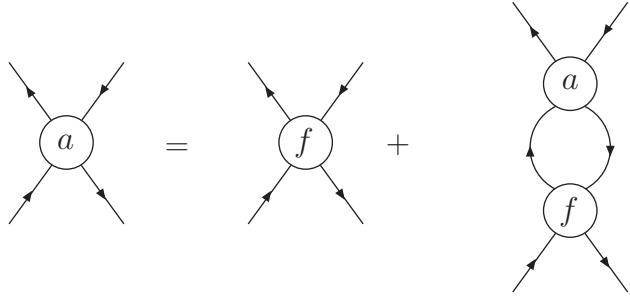}
\caption{The diagrammatic relationship between the physical scattering
  amplitude $a$ and the quasiparticle interaction $f$.} 
\label{Avsf}
\end{figure}

\begin{figure}
\includegraphics[height=3cm]{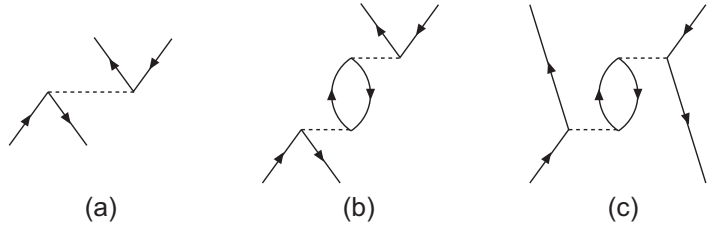}
\caption{Diagrams contributing to the quasiparticle interaction $f$ and the
  scattering amplitude $a$. Diagrams (a) and (c) contribute to $f$, whereas
  all three contribute to $a$.} 
\label{symm}
\end{figure}

In principle one could exactly calculate the quasiparticle interaction by
summing up all $ph$ irreducible diagrams contributing to the $ph$ vertex
function in the limit $k/\omega 
\rightarrow 0$. Since this is not practicable in general, one must limit the
calculation to a certain subset of diagrams. We could proceed by calculating
the relevant diagrams order by order, but this would miss an essential
point, which we now elaborate. From eqs.\ (\ref{scat}) and (\ref{fA}), we see
that the physical scattering amplitude 
$a(k_1,k_2)$ iterates the quasiparticle interaction to all orders through an
integral equation shown schematically in Fig.\ \ref{Avsf}. 
If only a finite set of diagrams are included in the quasiparticle
interaction, then the scattering amplitude will not be antisymmetric. For
instance, if we 
include only the bare particle-hole antisymmetrized vertex shown in Fig.\
\ref{symm}(a),
then diagram (b) will be contained in the equation for the
scattering amplitude but its exchange diagram, labeled (c), will not.
Quantitatively, the 
fact that the scattering amplitude is antisymmetric requires that it vanish
in singlet-odd and triplet-odd states as the Landau angle $\theta$ approaches
0. This leads to two constraints \cite{landau3, friman} on the Fermi liquid
parameters in the form of sum rules:
\begin{equation}
%\sum_l \left (\frac{F_l}{1+F_l/(2l+1)} +
%  \frac{F_l^\prime}{1+F_l^\prime/(2l+1)} +
%  \frac{G_l}{1+G_l/(2l+1)} +
%  \frac{G_l^\prime}{1+G_l^\prime/(2l+1)} \right ) =0
\sum_l \left (\frac{F_l}{1+F_l/(2l+1)} +
  3\frac{G_l^\prime}{1+G_l^\prime/(2l+1)} \right ) =0
\label{pauli1}
\end{equation}

\begin{equation}
\sum_l \left (\frac{2}{3}\frac{F_l}{1+F_l/(2l+1)} +
  \frac{F_l^\prime}{1+F_l^\prime/(2l+1)} +
  \frac{G_l}{1+G_l/(2l+1)} \right )=0
\label{pauli2}
\end{equation}

\begin{figure}
\includegraphics[height=4cm]{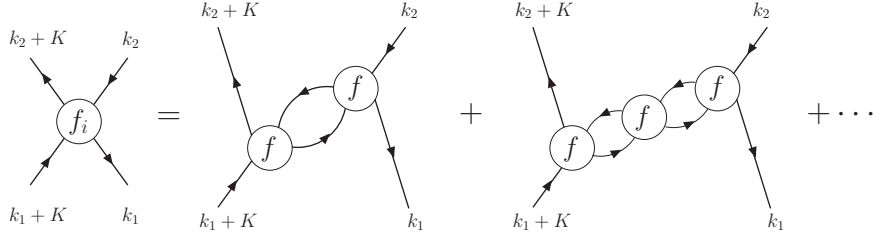}
\caption{The diagrammatic form of the induced interaction. In the limit that
  $k_1=k_2$ it can be shown that the external lines exactly couple to
  particle-hole excitations through the $f$ function.} 
\label{induced}
\end{figure}

Clearly, the sum rules must be satisfied for the ``correct'' set of Fermi
liquid parameters describing nuclear matter. To account for this
infinite set of exchange diagrams, Babu and Brown \cite{babu} proposed
separating the 
quasiparticle interaction into a {\it driving term} and an {\it induced term}:
\begin{equation}
f(k,k^\prime)=f_d(k,k^\prime) + f_i(k,k^\prime),
\label{split}
\end{equation}
where the induced interaction is defined to contain those diagrams that would
be the exchange terms necessary to preserve the antisymmetry of
$a(k_1,k_2)$. Then the induced interaction is
given by a diagrammatic expansion shown in Fig.\ \ref{induced}. Physically,
the induced interaction represents that part of the quasiparticle interaction
that results from the exchange of virtual collective modes, which can be
classified as density, spin, or isospin excitations.
In the limit that ${\bf k}_1
\rightarrow {\bf k}_2$ it can be rigorously proved \cite{babu} that the
coupling of quasiparticles to these collective excitations is precisely
through the quasiparticle interaction itself, thereby justifying the
diagrammatic expression in Fig.\ \ref{induced}.

The relationship between the induced interaction and the full quasiparticle
interaction was derived by Babu and Brown \cite{babu} for liquid $^3$He and
applied to nuclear matter by Sj\"{o}berg \cite{sjoberg1}. To lowest
order in the Fermi liquid parameters, the induced interaction is given by
\begin{eqnarray}
4F_i &=& \left[
          \frac{{F_0}^2}{1 + F_0\alpha_0}
       +  \frac{3{{F_0}^\prime}^2}{1 + {F_0}^\prime \alpha_0}
       +  \frac{{3G_0}^2}{1 + G_0\alpha_0}
       +  \frac{9{{G_0}^\prime}^2}{1 + {G_0}^\prime \alpha_0}\right]\alpha_0
\nonumber \\
4G_i &=& \left[ \frac{{F_0}^2}{1 + F_0\alpha_0}
       + \frac{{3{F_0}^\prime}^2}{1 + {F_0}^\prime \alpha_0}
       - \frac{{G_0}^2}{1 + G_0\alpha_0}
       - \frac{3{{G_0}^\prime}^2}{1 + {G_0}^\prime \alpha_0}\right]\alpha_0
\nonumber \\
4{F_i}^\prime &=& \left[ 
          \frac{{F_0}^2}{1 + F_0\alpha_0}
       - \frac{{{F_0}^\prime}^2}{1 + {F_0}^\prime \alpha_0}
       + \frac{{3G_0}^2}{1 + G_0\alpha_0}
       - \frac{3{{G_0}^\prime}^2}{1 + {G_0}^\prime \alpha_0}\right]\alpha_0
\nonumber \\
4{G_i}^\prime &=& \left[ \frac{{F_0}^2}{1 + F_0\alpha_0}
       - \frac{{{F_0}^\prime}^2}{1 + {F_0}^\prime \alpha_0}
       - \frac{{G_0}^2}{1 + G_0\alpha_0}
       + \frac{{{G_0}^\prime}^2}{1 + {G_0}^\prime \alpha_0}\right]\alpha_0
\label{first}
\end{eqnarray}
where 
\begin{equation}
\alpha_0=\alpha_0({\bf q},0)=\frac{1}{2}+\frac{1}{2}\left(
  \frac{q}{4k_F}-\frac{k_F}{q}\right ) {\rm ln}\frac{k_F-q/2}{k_F+q/2}
\end{equation}
is the Lindhard function, which is related to the density-density correlation
function $\chi_{\rho \rho}$ by 
\begin{equation}
\chi_{\rho \rho}({\bf q},\omega)=\frac{-\alpha_0({\bf
    q},\omega)}{1+F_0\alpha_0({\bf q},\omega)},
\end{equation}
and ${\bf q}={\bf k}_1-{\bf k}_2$. The interpretation of equation
(\ref{first}) is as 
follows. The Landau parameters in the numerator describe the coupling of
quasiparticles to particular collective modes. For instance, the $F_0$
represents the coupling to density excitations, $G_0$ the coupling
to spin excitations, etc., and the denominators enter from the summation of
bubbles to all orders. Including the $l=1$ Fermi liquid parameters, the induced
interaction is given by
\begin{eqnarray}
4F_i &=& \left[
         \frac{{F_0}^2\alpha_0}{1 + F_0\alpha_0}+\left(1 - \frac{q^2}{4k_F^2}
         \right ) \frac{{F_1}^2\alpha_1}{1 + F_1\alpha_1}\right. 
     + \frac{{3G_0}^2\alpha_0}{1 + G_0\alpha_0} + \left( 1-\frac{q^2}{4k_F^2}
         \right ) \frac{{3G_1}^2\alpha_1}{1 + G_1\alpha_1} \nonumber \\
    & +& \frac{3{{F_0}^\prime}^2\alpha_0}{1 + {F_0}^\prime \alpha_0}+\left(
         1-\frac{q^2}{4k_F^2}\right ) \frac{3{{F_1}^\prime}^2\alpha_1}{1 +
         {F_1}^\prime \alpha_1} + \left. \frac{9{{G_0}^\prime}^2\alpha_0}{1 +
         {G_0}^\prime \alpha_0} + \left( 1-\frac{q^2}{4k_F^2}\right )
         \frac{9{{G_1}^\prime}^2\alpha_1}{1 + {G_1}^\prime \alpha_1} \right ],
\label{second}
\end{eqnarray}
where $\alpha_1$ defined by 
\begin{equation}
\alpha_1({\bf q},0) = \frac{1}{2} \left [ \frac{3}{8}-\frac{k_F^2}{2q^2} +
  \left( \frac{k_F^3}{2q^3} + \frac{k_F}{4q} - \frac{3q}{32k_F} \right) {\rm
  ln} \left( \frac{k_F + q/2}{k_F - q/2}\right ) \right]
\end{equation}
is related to the current-current correlation function, and analogous
expressions hold for the spin- and isospin-dependent parts of the induced
interaction. These equations were first obtained in \cite{backman2},
carried far enough to include velocity-dependent effects in terms of an
effective mass, in the approximation of quadratic spectrum.

\begin{figure}
\includegraphics[height=3cm]{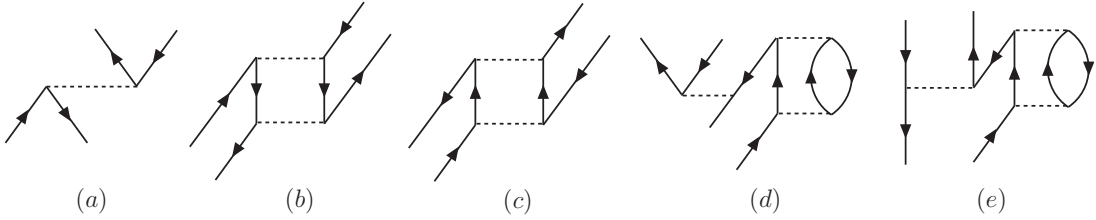}
\caption{A selection of diagrams contributing to the driving term in the
  quasiparticle interaction. Diagrams (d) and (e) are included implicitly
  through the renormalization at the quasiparticle pole.} 
\label{driving}
\end{figure}

Having characterized the induced part of the quasiparticle interaction, let us 
now elaborate on the driving term. By definition, this component of the
interaction consists of those diagrams that cannot be separated into two
diagrams by cutting one particle line and one hole line. Some of the low order
terms contributing to the driving term are shown in Fig.\ \ref{driving}, where
the interaction vertices are assumed to be antisymmetrized.
Some higher-order terms, such as diagram $(d)$ in Fig.\ \ref{driving}, are
included implicitly through the quasiparticle renormalization $z$ and need not
be calculated explicitly, as described in detail in \cite{brown2}. In order to
preserve the Pauli principle sum rules 
(\ref{pauli1}) and (\ref{pauli2}) the driving term must be
antisymmetrized. Thus, including Fig.\ \ref{driving}(d) requires that (e) also
be included in order for the scattering amplitude to be antisymmetric.

\section{Calculations and Results}

According to the discussion in the previous section, the starting point of a
microscopic derivation of the quasiparticle interaction is a calculation of
the antisymmetrized driving term to some specified order in the bare
potential. Nearly all previous calculations have used the $G$-matrix, since it
is well known that the unrenormalized high-precision NN potentials are
unsuitable 
for perturbation theory calculations due to the presence of a strong
short-distance repulsion. The resummation of particle-particle ladder diagrams
in the $G$-matrix softens the potential but introduces several undesirable
features from the perspective of Fermi liquid theory. Most important is the
unphysical gap in the single particle energy spectrum at the Fermi surface
due to the fact that hole lines receive self-energy corrections but particle
lines do not. In the past it was suggested \cite{poggioli,sjoberg2} that
introducing a model space, 
within which particles and holes are treated symmetrically, could overcome
this difficulty.

An alternative method for taming the repulsive core is to integrate out the
high momentum components of the interaction in such a way that the low energy
dynamics are preserved \cite{bogner1, bogner3}. This is accomplished by
rewriting the half-on-shell $T$-matrix
\begin{equation}
T(p^\prime,p,p^2) = V_{NN}(p^\prime,p) + \frac{2}{\pi}{\cal P} \int _0
^{\infty} \frac{V_{NN}(p^\prime,q)T(q,p,p^2)} {p^2-q^2} q^2 dq
\end{equation}
with an explicit momentum cutoff $\Lambda$, which yields the low momentum
$T$-matrix defined by
\begin{equation}
T_{{\rm low}-k }(p^\prime,p,p^2) = V_{{\rm low}-k}(p^\prime,p) +
\frac{2}{\pi}{\cal P} \int _0 ^{\Lambda} \frac{V_{{\rm low}-k }(p^\prime,q)
  T_{{\rm low}-k} (q,p,p^2)}{p^2-q^2} q^2 dq.
\end{equation}
Enforcing the requirement that $T_{{\rm low}-k}(p^\prime,p,p^2) =
T(p^\prime,p,p^2)$ for $p^\prime, p < \Lambda$ preserves the low energy
physics encoded in the scattering phase shifts. Remarkably, under
this construction all high-precision NN potentials flow to a nearly universal
low momentum interaction \vlk as the momentum cutoff $\Lambda$ is lowered to
$2.1$ fm$^{-1}$. In fact, $k=2.1$ fm$^{-1}$ is precisely the CM 
momentum beyond which the experimental phase shift analysis has not been
incorporated in the high-precision NN interactions.

For an initial approximation to the driving term, we include the first-order
antisymmetrized matrix element shown diagrammatically in Fig.\
\ref{driving}(a) as well  
as the higher order diagrams, such as (d) and (e), that are included
implicitly through the renormalization strength at the quasiparticle
pole. The quasiparticles are confined to a thin model space $P$ near the Fermi
surface 
\begin{equation}
P=\lim_{\delta \to 0} \sum_{k_F<k<k_F+\delta} | \vec{k} \rangle \langle
\vec{k} |,
\end{equation}
and the first-order contribution is given by
\begin{equation}
\langle \vec{k}_1\vec{k}_2 ST| V | (\vec{k}_3\vec{k}_4 - \vec{k}_4\vec{k}_3)
ST\rangle = \langle k,\theta ST |V| k, \theta ST \rangle,
\end{equation}
where $k_1=k_2=k_3=k_4=k_F$, $\theta$ is the angle between the two momenta, and
the relative momentum $k=k_F \mbox{sin}(\theta /2)$. Given the \vlk matrix
elements in the basis $|klSTJ\rangle$, we project onto the central
components and change from a spherical wave basis to a plane wave basis. Then
the dimensionful driving term is given by
\begin{equation}
\langle kST|V_d|kST \rangle = z^2\frac{4\pi}{2S+1} \sum_{J,l} (2J+1)
(1-(-1)^{l+S+T}) \langle klSJT| V_{{\rm low-}k} | klSJT \rangle.
\label{me}
\end{equation}
Inserting the form of the quasiparticle
interaction in eq.\ (\ref{ffunction}) into the left hand side of 
eq.\ (\ref{me}), we obtain the Fermi liquid parameters in terms
of $V_{ST}(k) = \langle kST|V|kST\rangle$. The result is
\begin{eqnarray}
f &=&\; \; \, \frac{1}{16}V_{00} + \frac{3}{16}V_{01} + \frac{3}{16}V_{10} +
\frac{9}{16}V_{11} \nonumber \\
g &=& -\frac{1}{16}V_{00} - \frac{3}{16}V_{01} + \frac{1}{16}V_{10} +
\frac{3}{16}V_{11} \nonumber \\
f^\prime &=& -\frac{1}{16}V_{00} + \frac{1}{16}V_{01} - \frac{3}{16}V_{10} +
\frac{3}{16}V_{11} \nonumber \\
g^\prime &=& \; \; \, \frac{1}{16}V_{00} - \frac{1}{16}V_{01} -
\frac{1}{16}V_{10} + \frac{1}{16}V_{11},
\end{eqnarray}
where the momentum dependence has been suppressed for simplicity. From eq.\
(\ref{effmass}) it can be shown that 
\begin{equation}
\frac{m^*}{m}=\frac{1}{1-\mu f_1/3},
\end{equation}
where $\mu=2mk_F/\pi^2 \hbar^2=\frac{m}{m^*}N_0$, from which we construct
the dimensionless Fermi 
liquid parameters. In all of our calculations we include partial waves up to
$J=6$. In Table \ref{drivetab} 
we show the Landau parameters of the driving term derived from three different
low momentum interactions obtained from the Nijmegen I \& II potentials
\cite{wiringa} and the CD-Bonn potential \cite{machleidt} for a momentum
cutoff of $\Lambda =2.1$ fm$^{-1}$ and a Fermi momentum of $k_F=1.36$
fm$^{-1}$. From the available theoretical analyses of nucleon momentum
distributions \cite{pand}, we take the quasiparticle renormalization strength
to be $z = 0.7$ for nuclear matter.
\setlength{\tabcolsep}{.075in}
\begin{table}
\begin{tabular}{|c|c|c|c|c||c|c|c|c||c|c|c|c|} \hline
\multicolumn{1}{|c|}{} & \multicolumn{4}{|c||}{Nijmegen I} &
\multicolumn{4}{|c||}{Nijmegen II} & \multicolumn{4}{|c|}{CD-Bonn} \\ \hline
$l$ & $F_l$ & $G_l$ & $F^\prime_l$ & $G^\prime_l$ & $F_l$ & $G_l$ &
$F^\prime_l$ & $G^\prime_l$ & $F_l$ & $G_l$ & $F^\prime_l$ & $G^\prime_l$  \\
\hline 

% z^2 error not corrected in argonne data
%0 & -1.646 & 0.194 & 0.491 & 0.784 & -1.490 & 0.168 & 0.436 & 0.750 \\ \hline
%1 & -0.653 & 0.307 & 0.359 & 0.174 & -0.619 & 0.299 & 0.321 & 0.146 \\ \hline
%2 & -0.271 & 0.149 & 0.143 & 0.036 & -0.248 & 0.151 & 0.125 & 0.028 \\ \hline
%3 & -0.150 & 0.070 & 0.069 & 0.014 & -0.138 & 0.069 & 0.063 & 0.013 \\ \hline

% Nijmegen II & CD-Bonn   ! (z=0.7)
0 & -1.230 & 0.130 & 0.392 & 0.619 & -1.475 & 0.248 & 0.549 & 0.583 & -1.199 &
0.135 & 0.350 & 0.603 \\ \hline 
1 & -0.506 & 0.241 & 0.252 & 0.118 & -0.445 & 0.161 & 0.172 & 0.225 & -0.498 &
0.240 & 0.259 & 0.118 \\ \hline 
2 & -0.201 & 0.120 & 0.101 & 0.021 & -0.213 & 0.127 & 0.106 & 0.020 & -0.200 &
0.122 & 0.101 & 0.022 \\ \hline 
3 & -0.110 & 0.054 & 0.051 & 0.009 & -0.120 & 0.060 & 0.056 & 0.007 & -0.111 &
0.055 & 0.051 & 0.010 \\ \hline 

\end{tabular}
\caption{The Fermi liquid parameters of the NN interaction \vlk derived from
the Nijmegen potentials and CD-Bonn potential for a cutoff of $\Lambda = 2.1$
fm$^{-1}$ and Fermi momentum 1.36 fm$^{-1}$.}
\label{drivetab}
\end{table}

The induced interaction is obtained by iterating equations (\ref{split}) and
(\ref{second}) until a self-consistent solution is reached. The
density-density and current-current correlation functions in (\ref{second})
introduce a momentum dependence in the induced interaction, and the
Fermi liquid parameters for the induced interaction are obtained by projecting
onto the Legendre polynomials
\begin{equation}
F_{i,l}=\frac{2l+1}{2} \int_{-1}^1 F_i(\theta) P_l(\mbox{cos }\theta)
\, d(\mbox{cos }\theta), \mbox{ etc.}
\end{equation}
For the first iteration we use the Landau parameters obtained from the bare
low momentum interaction as an estimate for 
the full quasiparticle interaction in eq.\ (\ref{second}). However, since
$F_0$ does not satisfy the stability criteria (\ref{stab}) for either the
Nijmegen or CD-Bonn potentials, in the first iteration we replace it in both
cases with an arbitrary value that does. The convergence of
the iteration scheme is generally rapid and relatively insensitive to the set
of initial parameters chosen for the low-momentum Nijmegen I and CD-Bonn
potentials. In contrast, the low momentum Nijmegen II potential exhibits poor
convergence properties, though a solution to the coupled equations can still be
found. For a completely consistent solution at each iteration we recalculate
the driving term with the new effective mass.

\begin{figure}[tbh]
\includegraphics[height=15cm,angle=-90]{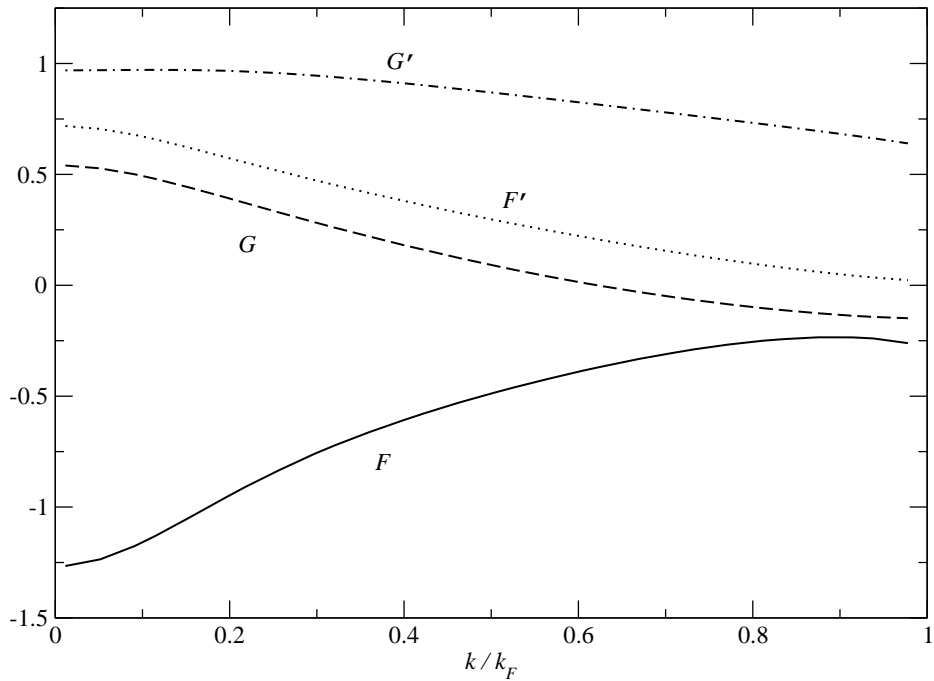}
\caption{The self-consistent solution for the full quasiparticle interaction
  as a function of $k=\frac{1}{2}|{\bf k}_1 - {\bf k}_2|$ derived from
  the low momentum CD-Bonn potential.}
\label{FLP}
\end{figure}

The final self-consistent result for the quasiparticle interaction is
shown in Fig.\ \ref{FLP} for the CD-Bonn potential, and the Fermi liquid
parameters for the driving term, induced interaction, and full quasiparticle
interaction are shown in Table \ref{fulltab}. 
\setlength{\tabcolsep}{.07in}
\begin{table}
\begin{tabular}{|c||c|c|c|c||c|c|c|c||c|c|c|c|} \hline
\multicolumn{1}{|c||}{} & \multicolumn{4}{|c||}{Full} &
\multicolumn{4}{|c||}{Driving} & \multicolumn{4}{|c|}{Induced} \\ \hline
$l$ & $F$ & $G$ & $F^\prime$ & $G^\prime$ & $F_d$ & $G_d$ &
$F^\prime_d$ & $G^\prime_d$ & $F_i$ & $G_i$ & $F^\prime_i$ & $G^\prime_i$ \\
\hline 
0 & -0.476 & 0.025 & 0.221 & 0.784 & -1.276 & 0.144 & 0.373 & 0.642 
& 0.801 & -0.119 & -0.152 & 0.142 \\ \hline
1 & -0.335 & 0.263 & 0.273 & 0.171 & -0.530 & 0.256 & 0.275 & 0.125 
& 0.195 & 0.007 & -0.002 & 0.048 \\ \hline
2 & -0.238 & 0.139 & 0.117 & 0.020 & -0.212 & 0.130 & 0.107 & 0.024 
& -0.026 & 0.009 & 0.010 & -0.003 \\ \hline
3 & -0.101 & 0.055 & 0.050 & 0.014 & -0.119 & 0.059 & 0.054 & 0.011 
& 0.018 & -0.004 & -0.004 & 0.003 \\ \hline
\end{tabular}
\caption{The self-consistent solution of the Babu-Brown equations for the low
  momentum CD-Bonn potential. The full
  Fermi liquid parameters are obtained by projecting the quasiparticle
  interaction in Fig.\ \ref{FLP} onto the Legendre polynomials.}
\label{fulltab}
\end{table}
For comparison we list the Fermi liquid parameters
obtained in \cite{schwenk1} where the spin-independent Landau parameters were
taken from experiment and used to calculate the spin-dependent parameters
with a set of nontrivial sum rules:
\begin{center}
\setlength{\tabcolsep}{.3in}
\begin{tabular}{llll}
$F_0 = -0.27$ & $G_0 = 0.15 \pm 0.3$ & ${F_0}^\prime = 0.71$ & ${G_0}^\prime =
1.0 \pm 0.2$ \\
$F_1 = -0.85$ & $G_1 = 0.45 \pm 0.3$ & ${F_1}^\prime = 0.14$ & ${G_1}^\prime =
0.0 \pm 0.2$ 
\end{tabular}
\end{center}
Although the experimental values for the spin-independent parameters are
appreciably different from the self-consistent solution we have obtained,
our values for the spin-dependent parameters fall within the errors predicted
from the sum rules. However, the main effect of the induced interaction is to
cut down the strong attraction in the spin-independent, isospin-independent
part of the quasiparticle interaction. In fact, the repulsion in this channel
coming from the induced interaction is large enough for the resulting $F_0$ to
satisfy the stability condition in (\ref{stab}). The effective mass,
compression modulus, and symmetry energy are shown in Table \ref{fintab}
together with the deviations $\delta S_1$ and $\delta S_2$ from the sum rules
(\ref{pauli1}) 
and (\ref{pauli2}). We list the results for the three different bare potentials
with a momentum cutoff of $\Lambda = 2.1$ fm$^{-1}$. In calculating the
contributions to (\ref{pauli1}) and (\ref{pauli2}) we have included Landau
parameters for $l \leq 3$. 
\setlength{\tabcolsep}{.1in}
\begin{table}
\begin{tabular}{|c||c|c|c|} \hline

% z^2 not corrected here
% & Argonne $v_{18}$ & CD-Bonn \\ \hline  %z=0.8
%$m^*/m$ & 0.859 & 0.861  \\ \hline 
%K [MeV]&  128 & 137  \\ \hline
%$\beta$ [MeV] & 18.8 & 18.2  \\ \hline 
%$\delta S_1$ & 0.044 & 0.054  \\ \hline 
%$\delta S_2$ & 1.17 & 1.18  \\ \hline

& Nijmegen I & Nijmegen II & CD-Bonn \\ \hline
$m^*/m$ & 0.887 & 0.930 & 0.888  \\ \hline   %z=0.7
${\cal K}$ [MeV]&  136 & 102 & 136  \\ \hline
$\beta$ [MeV] & 18.1 & 20.5 & 17.6  \\ \hline
$\delta g_l$ [$\mu_N$] & 0.682 & 0.452 & 0.685 \\ \hline
$\delta S_1$ & 0.20 & 0.16 & 0.27  \\ \hline
$\delta S_2$ & -0.04 & -0.02 & -0.04  \\ \hline

\end{tabular}
\caption{Nuclear observables obtained from the self-consistent solution of the
  Babu-Brown equations and deviations $\delta S_1$ and $\delta S_2$ from the
  Pauli principle sum rules.}
\label{fintab}
\end{table}
The compression modulus for nuclear matter is extrapolated from the data on
giant monopole resonances in heavy nuclei, with the expected value being 
$200-300$ MeV \cite{young,steiner}. The symmetry energy is determined by
fitting the data on nuclear masses to various versions of the semi-empirical
mass formula \cite{masses}, and currently the accepted value is $\beta = 25
\mbox{--} 35$ MeV \cite{danielewicz,steiner}. Both the compression modulus and
the symmetry energy shown in Table \ref{fintab} are significantly smaller than
the experimental values. On the other hand, the anomalous orbital gyromagnetic
ratio, determined from giant dipole resonances in heavy nuclei, is too large
compared with the experimental value of $\delta g_l^p = 0.23 \pm
0.03$ \cite{nolte}.

\begin{figure}
\includegraphics[height=5cm]{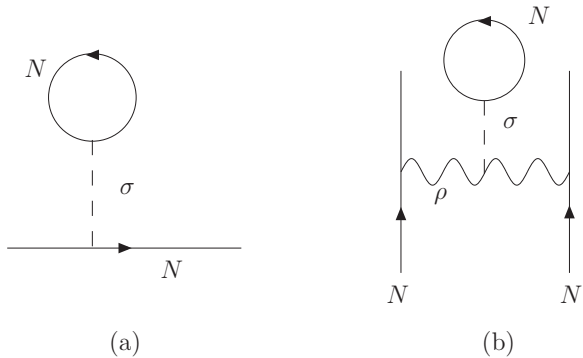}
\caption{Walecka mean field contribution of the scalar tadpole to the
  nucleon mass (a) and its extrapolation to constituent quarks in vector
  mesons (b).}
\label{scalar}
\end{figure}

As suggested in the introduction, we propose to remedy these discrepancies by
considering the effects of Brown-Rho scaling on hadronic masses. The proposed
scaling law for light hadrons -- other than the pseudoscalar mesons, whose
masses are protected by chiral invariance -- is \cite{brownrho,brownpr}
\begin{equation}
\frac{m_V^*}{m_V} = \frac{m_\sigma^*}{m_\sigma} = 
\sqrt{\frac{g_A}{g_A^*}}\frac{m_N^*}{m_N} = 1-C \frac{n}{n_0},
\label{escale}
\end{equation}
where the subscript $V$ denotes either the $\rho$ or $\omega$ vector meson,
$\sigma$ refers to the scalar meson, $g_A$ is the axial vector coupling, and
$n/n_0$ is the ratio of the medium 
density to nuclear matter density. This scaling can be thought of as extending
Walecka mean field theory, in which the scalar tadpole contribution 
to the nucleon self-energy lowers the effective mass, to the level of
constituent quarks. Attaching a scalar tadpole on the nucleon line, as shown in
Fig.\ \ref{scalar}(a), lowers the mass according to (\ref{escale}), and a
scalar tadpole connected to the vector mesons gives an effective three-body
force as shown in Fig.\ \ref{scalar}(b). Including the in-medium scaling of the
axial-vector coupling, which should approach $g_A^*
= 1$ at chiral restoration, the net result is a lowering of the in-medium
$m_V^*$ by $\sim 2/3$  
as much as $m_N^*$. Recent experimental results \cite{trnka, naruki} are
consistent with the scaling law (\ref{escale}) for $C = 0.15$ and
$0.092$, respectively. The Brown-Rho ``parametric scaling'' has
$C=0.2$. However, the dense loop term $\Delta M$ \cite{hkr} gives a
shift of the $\rho$-meson pole upwards. So far no one has been able to
calculate it at finite density.

A number of previous studies \cite{friman2,song1,song2} were successful in
describing nuclear matter by starting from a chiral Lagrangian with nuclear,
scalar, and vector degrees of freedom in which the hadronic masses were scaled
with density according to (\ref{escale}). In particular, the compression
modulus and anomalous orbital gyromagnetic ratio were found to be in excellent
agreement with experiment, which suggests that a similar
approach may prove fruitful in our present analysis. An alternative approach,
complementary to the chiral Lagrangian method, is to include medium
modifications directly into a one-boson-exchange potential. Such a calculation
was carried out in \cite{rapp} to study the saturation of nuclear matter. In
their work it was suggested that the $\sigma$ particle should be constructed
microscopically as a pair of correlated pions interacting largely through
crossed-channel $\rho$ exchange. Medium modifications to the $\sigma$ mass
then arises naturally from the density-dependence of the $\rho$ mass. The
final conclusion established in \cite{rapp} is that at low densities the
$\sigma$ scales according to (\ref{escale}) but that toward nuclear matter
density the scaling is slowed to such an extent that saturation can be
achieved.

We proceed along the lines of \cite{rapp} and introduce medium modifications
directly into a one-boson-exchange potential. The most refined NN
potentials in this category are the Nijmegen I, Nijmegen II, and CD-Bonn
potentials. The Nijmegen potentials include contributions
from the exchange of $\rho, \omega, \phi, \sigma, f_0,$ and $a_0$ mesons, as
well as the pseudoscalar particles which do not receive medium modifications
in Brown-Rho scaling. The CD-Bonn potential includes two vector particles (the
$\rho$ and $\omega$) and two scalars ($\sigma_1$ and $\sigma_2$). For both
potentials we scale the vector meson masses by 15\% and the scalar meson
masses by 7\%. In this way we roughly account for the decreased scaling of the 
scalar particle mass observed in \cite{rapp}. In the full many-body calculation
we also scale the nucleon mass by 15\% and with an additional
$\sqrt{g_A^*/g_A} \simeq 1/\sqrt{1.25}$ at nuclear matter density. It is
essential to also scale the form factor cutoffs $\Lambda_f$ of the vector
mesons in the boson-exchange potentials.

In Table \ref{brtab} we show the effective mass, compression modulus, symmetry
energy, and anomalous orbital gyromagnetic ratio for the Nijmegen I \& II and
CD-Bonn potentials with the in-medium modifications. We also show for
comparison the results from the Nijmegen93 one-boson-exchange potential, which
has only 15 free parameters and is not fine-tuned separately in each partial
wave. We observe that the iterative solution is in better agreement with all
nuclear observables. The anomalously large compression modulus in the CD-Bonn
potential results almost completely from the presence of a large $\omega$
coupling constant $g_{\omega N N}^2/4\pi = 20.0$. With the same $g_{\omega N
  N}^2/4\pi$ and Bonn-B potential, Rapp {\it et al.} \cite{rapp} obtained
${\cal K} = 356$ MeV. The compression modulus is
very sensitive to this parameter, as we have checked that
dropping this coupling by 20\% cuts the compression modulus in half but alters
the other nuclear observables by less than 5\%. The naive quark model
predicts a ratio of $g_{\omega NN}^2/g_{\rho NN}^2 = 9$ between the $\omega$
and $\rho$ coupling constants, which is largely violated in the CD-Bonn
potential $g_{\omega NN}^2/g_{\rho NN}^2 = 24$ though roughly satisfied in the
Nijmegen potentials $g_{\omega NN}^2/g_{\rho NN}^2 = 11$, perhaps resulting in
better agreement with experiment. 

%These high precision models of the NN interaction
%achieve remarkably good agreement with scattering phase shift data
%($\chi^2$/DOF $\sim 1$) by fine tuning certain parameters separately in each
%partial wave. In both potentials the mass of the lightest scalar particle is
%one of these parameters and accounts for part of the 

Thus, by extension of the Walecka mean field on nucleons to those on
constituent quarks, we obtain the Fermi liquid parameters for the theory that
is now essentially Brown-Rho scaled, as shown in Table \ref{brfintab}. One
should note that these results are only for infinite nuclear matter and
especially the three-body term will act in many different diagrams in the
finite systems. However, our arguments suggest that the three-body terms
intrinsic to Brown-Rho scaling will be useful in stabilizing light nuclei.
%\setlength{\tabcolsep}{.1in}
%\begin{table}
%\begin{tabular}{|c|c|c|c|c||c|c|c|c|} \hline
%\multicolumn{1}{|c|}{CD-Bonn} & \multicolumn{2}{|c|}{$\frac{g^2_{\omega
%      NN}}{4\pi}=20.0$} & \multicolumn{2}{|c||}{$\frac{g^2_{\omega
%      NN}}{4\pi}=7.56$} & \multicolumn{2}{|c|}{Nijmegen I} &  
%\multicolumn{2}{|c|}{Nijmegen II} \\ \hline 
%$\lambda$ & 0.14 & 0.092 & 0.14 & 0.092 & 0.14 & 0.092 & 0.14 & 0.092 \\
%\hline 
%$m^*/m$ & 0.642 & 0.717 & 0.645 & 0.673 & 0.695 & .702 & 0.655 & 0.717\\
%\hline 
%K [MeV] & 586 & 404 & 223 & 164 & 262 & 207 & 357 & 272 \\ \hline
%$\beta$ [MeV] & 21.2 & 21.1 & 27.9 & 27.9 & 20.1 & 22.1 & 22.2 & 22.6\\
%\hline  
%\end{tabular}
%\caption{Nuclear observables obtained from the self-consistent solution to the
%  Babu-Brown equations incorporating Brown-Rho scaling. The parameter
%  $\lambda$ is defined in eq.\ (\ref{escale}).} 
%\label{brtab}
%\end{table}
\setlength{\tabcolsep}{.1in}
\begin{table}
\begin{tabular}{|c|c|c|c|c|} \hline
         & $V_{\rm NI}$  & $V_{\rm NII}$ & $V_{N93}$ & $V_{\rm CDB}$ \\ \hline 
$m^*/m$  & 0.721          & 0.763  &  0.696     & 0.682 \\ \hline
${\cal K}$ [MeV] & 218    & 142    &  190       & 495   \\ \hline
$\beta$ [MeV]    & 20.4   & 25.5   &  23.7      & 19.2  \\ \hline
$\delta g_l$     & 0.246  & 0.181  &  0.283     & 0.267 \\ \hline
\end{tabular}
\caption{Nuclear observables obtained from the self-consistent solution to the
  Babu-Brown equations incorporating Brown-Rho scaling. Four different bare
  potentials -- the CD-Bonn potential ($V_{\rm CDB}$), Nijmegen I ($V_{\rm
  NI}$), Nijmegen II ($V_{\rm NII}$), and Nijmegen93 ($N93$) potentials --
  were used to construct low momentum interactions for a cutoff of
  $\Lambda=2.1$ fm$^{-1}$. In eq.\ (\ref{escale}) the parameter $C = 0.15$.} 
\label{brtab}
\end{table}

\setlength{\tabcolsep}{.1in}
\begin{table}
\begin{tabular}{|c|c|c|c|c|} \hline
$l$ & $F_l$ & $G_l$ & $F^\prime_l$ & $G^\prime_l$   \\ \hline
0 & -0.20  $\pm$ 0.39  & 0.04  $\pm$ 0.11   & 0.24  $\pm$ 0.16
  &  0.53  $\pm$ 0.09 \\ \hline
1 & -0.86  $\pm$ 0.10  &  0.19  $\pm$ 0.06  & 0.18  $\pm$ 0.05
  &  0.17  $\pm$ 0.12 \\ \hline
2 & -0.21  $\pm$ 0.01  &  0.12 $\pm$ 0.01   & 0.10  $\pm$ 0.02  
  &  0.01  $\pm$ 0.02 \\ \hline
3 & -0.09  $\pm$ 0.01  &  0.05 $\pm$ 0.01  & 0.05   $\pm$ 0.01  
  &  0.01  $\pm$ 0.01 \\

%$l$ & $F_l$ & $G_l$ & $F^\prime_l$ & $G^\prime_l$   \\ \hline
%0 & -0.28  $\pm$ 0.04  & -0.01  $\pm$ 0.06  & 0.28  $\pm$ 0.05
%  &  0.64  $\pm$ 0.03 \\ \hline
%1 & -0.91  $\pm$ 0.03  &  0.22  $\pm$ 0.05  & 0.26  $\pm$ 0.04
%  &  0.14  $\pm$ 0.01 \\ \hline
%2 & -0.24  $\pm$ 0.02  &  0.12 $\pm$ 0.02  & 0.10  $\pm$ 0.02  
%  &  0.015 $\pm$ 0.008 \\ \hline
%3 & -0.094 $\pm$ 0.005 &  0.041 $\pm$ 0.002 & 0.042 $\pm$ 0.004  
%  &  0.009 $\pm$ 0.002 \\
\hline
\end{tabular}
\caption{Fermi liquid coefficients for the self-consistent solution to the
  Babu-Brown equations using Brown-Rho scaled nucleon and meson
  masses in the four low momentum CD-Bonn and Nijmegen potentials listed in
  Table \ref{brtab}. The tabulated values display the average and spread from
  the four different potentials and not the actual uncertainties associated
  with the Fermi liquid parameters.}
\label{brfintab}
\end{table}

\section{Discussion of the tensor force with dropping $\rho$-mass in
  saturation of nuclear matter}

The tensor force contributes chiefly in second order perturbation theory as an
effective central force in the $I=1$ channel. As the density increases, some
of the intermediate states are blocked by the
Pauli principle. In the two-body system the tensor force contributes to the
$^3S_1$ state, but not to the $^1S_0$ state, and gives most of the attractive
interaction difference between the $^3S_1$ and $^1S_0$ states, effectively
binding 
the deuteron. However, the intermediate state energies relevant for the
second-order tensor force are $> 225$ MeV (See Fig.\ 69 of \cite{brown3} which
is for $^{40}$Ca. For nuclear matter the intermediate state momenta would be
higher.),
well above the Fermi energy of nuclear matter, and most intermediate momenta
are above the \vlk upper model space limit of 420 MeV/c, so the tensor force
is largely integrated out.

\begin{figure}
\includegraphics[height=14cm, angle=-90]{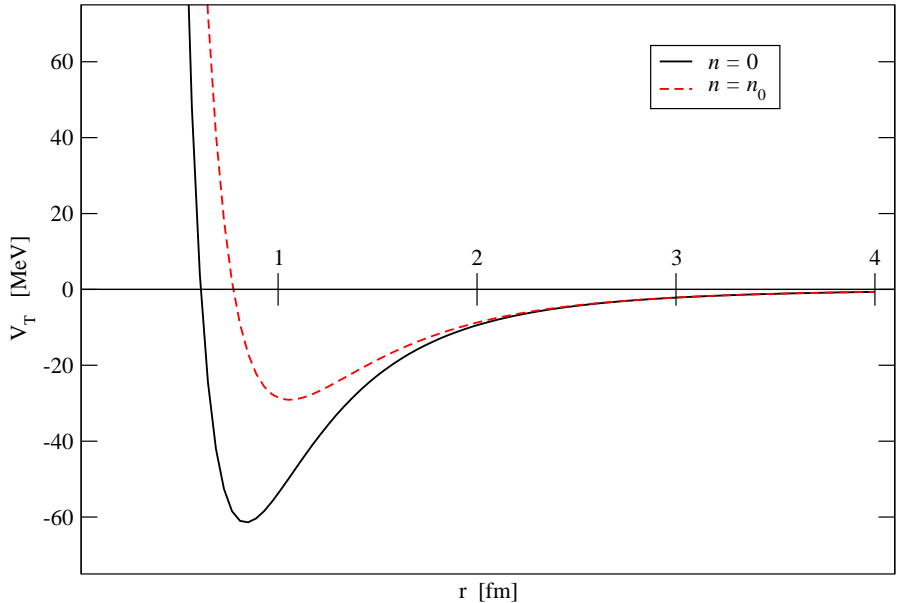}
\caption{Reduction in the strength of the tensor force due to a scaled
  $\rho$-meson mass. Contributions from both $\pi$-meson and $\rho$-meson
  exchange are included in both curves. We have used the Brown-Rho parametric
  scaling, so that at nuclear matter density $m^*_\rho = 0.8 m_\rho$.}
\label{tens}
\end{figure}

\begin{figure}
\includegraphics[height=14cm, angle=-90]{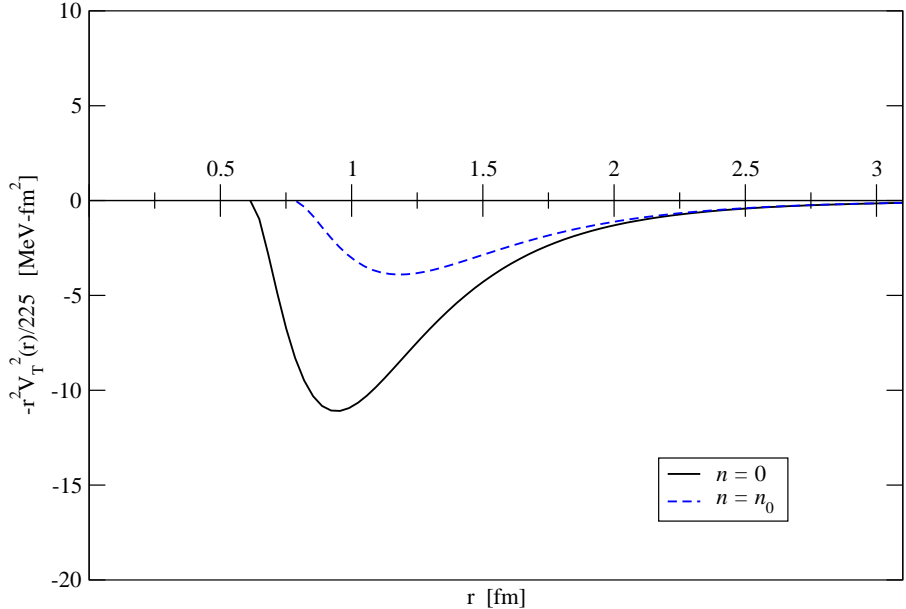}
\caption{Reduction of the tensor force in second order perturbation
  theory due to a scaled $\rho$-meson mass. The  
  intermediate state energy is approximated as 225 MeV. Contributions
  from both $\pi$-meson and $\rho$-meson exchange are included in both
  curves. At nuclear matter density, $n_0$, we have used the parametric
  scaling $m^*_\rho = 0.8 m_\rho$.}
\label{stens}
\end{figure}

However, since the beginning of Brown-Rho scaling it has been understood that
the tensor force is rapidly cut down with increasing density. That is because
the pion mass does not change with density, being protected by chiral
invariance, but the $\rho$-meson mass, which is dynamically generated,
decreases by 20\% (parametric scaling) in going from a density of $n=0$ to
nuclear matter density $n=n_0$. Since the 
$\rho$-meson exchange contributes with opposite sign from that of the pion,
this cuts down the tensor force substantially. In Fig.\ \ref{tens} we show the
total tensor force from $\pi$ and $\rho$ exchange at zero density and nuclear
matter density $n_0$. Since it enters in the square, this means a factor of
several drop in the tensor contribution to the binding energy, as shown in
Fig.\ \ref{stens}.

We believe that the work of ref.\ \cite{trnka} shows unambiguously that the
mass of the $\omega$-meson is $\sim$ 14\% lower at nuclear matter density than
in free space. It is remarkable that nuclear structure calculations have been
carried out for many years without density-dependent masses but with results
usually in quantitative agreement with experiment. In \cite{brown5} Brown
and Rho showed that in cases where the exchange of the $\pi$-meson is not
important, such as in Dirac phenomenology, there is a scale invariance such
that if the masses of all relevant mesons are changed by the same amount, the
results for the physical phenomena are very little changed.

Since the pion exchange gives the longest range part of the nucleon-nucleon
interaction, it is amazing that there are not clearcut examples in nuclear
spectroscopy such as level orderings that are altered by the $\rho$-meson
exchange playing counterpoint to the $\pi$-meson exchange, as we find in this
paper for nuclear saturation. The in-medium decrease in the $\rho$-mass
increases the effect of $\rho$-exchange, which enters so as to cut down the
overall tensor force, the $\rho$ and $\pi$ exchange entering with opposite
sign.

Finally, nearly forty years since the Kuo-Brown nucleon-nucleon forces were
first published, it was shown \cite{holt1} that the summation of core
polarization diagrams to all orders is well-approximated by a single
bubble. However, in light of the double 
decimation of \cite{brown5} being carried out here in one step, these forces
should be modified to include the medium dependence of the
masses. Phenomenologically this can be done by introducing three-body terms,
as we did here, but from our treatment of the second-order tensor force it is
clear that this should be done at constituent quark level.

\section{Conclusion}
We believe that by discussing the nuclear many-body problem within the context
of Fermi liquid theory with the interaction \vlk following the work of Schwenk
{\it et al.} \cite{schwenk1} we have a format for understanding connections
between the physical properties of the many-body system and the nuclear
potentials. We carried out an iterative solution of the Babu-Brown equations,
which include both density-density and current-current correlation functions,
calculating input potentials via a momentum space decimation to \vlkn. By
including Brown-Rho scaling through scalar tadpoles, as suggested by Walecka
theory, our iterative solution provides the empirical Fermi liquid
quantities. Our nucleon effective mass is on the low side of those usually
employed, as is common in Walecka mean field theory.

\begin{acknowledgements}
We thank Achim Schwenk for helpful discussions. This work was partially
supported by the U.\ S.\ Department of Energy under Grant
No.\ DE-FG02-88ER40388. 
\end{acknowledgements}

\end{document}